# Packet-wise multi-hop quantum routing of hybrid quantum-classical packets

TAE YEONG PARK,[1,2] YOUN CHANG JEONG,[3] MYEONG SOO KANG,[2,4]
KWANG YONG SONG,[5] PAUL G. KWIAT,[6] AND HEE SU PARK[1,2,*]

[1]*Quantum Technology Institute, Korea Research Institute of Standards and Science (KRISS), Daejeon 34113, South Korea*
[2]*Graduate School of Quantum Science, Korea Advanced Institute of Science and Technology (KAIST), Daejeon 34141, South Korea*
[3]*The Affiliated Institute of Electronics and Telecommunications Research Institute, Daejeon 34044, South Korea*
[4]*Department of Physics, KAIST, Daejeon 34141, South Korea*
[5]*Department of Physics, Chung-Ang University, Seoul 06974, South Korea*
[6]*Department of Physics, University of Illinois at Urbana-Champaign, Urbana, IL 61801, USA*
**hspark@kriss.re.kr*

**Abstract:** The ability to dynamically route quantum information across multiple nodes is a key requirement for realizing versatile and scalable quantum networks. Quantum routers are devices to achieve packet-level routing beyond static point-to-point connections by handling hybrid packets combining classical headers and quantum payloads. Previously proposed methods have selectively accessed the classical and quantum components of the packets utilizing wavelength splitters and optical switches, and repacked the hybrid packets before transmission through the outgoing link with regenerating classical headers. This work reports a novel optical-fiber-based quantum router that minimizes the disturbance and latency on the hybrid packets through more transparent operation. We experimentally demonstrate double-hop routing via two cascaded routers to three distinct receivers without regeneration of the headers. As a practical application, quantum key distribution packets based on time-bin qubits and a BB84 protocol with a two-decoy-state method are successfully delivered to the three receivers achieving nonzero secret key rates. These results demonstrate the feasibility of extensive quantum networks interconnected by multiple-step router nodes.

## 1. Introduction

The development of quantum networks is advancing towards realization of reconfigurable platforms capable of connecting multiple quantum information processing units. In the short term, such infrastructures enable cryptographic communication tasks with provable security based on quantum information theories [1–3], and can further implement transparent quantum information links for distributed quantum computing [4–6] and sensing [7–9] to step up the scale of quantum information processing. Quantum routing in the current work aims to maximize the utility and efficiency of network resources by realizing dynamical packet-wise connections between quantum nodes [10,11]. A quantum router sets the paths of incoming quantum information packets based on destination address tags, analogous to those used in classical communication systems. This inevitably requires the use of hybrid packets comprising classical headers and quantum payloads alongside a routing capability that can keep quantum information intact during transmission.

Previously multi-party quantum networks have mostly been implemented using wavelength-division multiplexing (WDM) passive architectures [12–14] or optical switch-based configurations [15,16]. WDM-based techniques allocate predefined wavelengths to nodes to distribute quantum information to specific channels by wavelength-selective elements. Use of multiple optical switches for the latter cases creates a direct connection between a sender and a receiver with a universally configurable optical switch matrix that scales according to the number of nodes. In parallel with the optical hardware, quantum software-defined networks (QSDNs) have been proposed to logically integrate classical and quantum networks for configuring quantum links across the entire network [17–21]. The QSDNs were incorporated to operate heterogeneous quantum key distribution (QKD) equipment across multiple sites, carrier domains, and services. However, such combination of conventional architectures and a QSDN require global control over the switching/branching elements to schedule all the sender-receiver pair instances. Additionally, integration of separately operated quantum networks as a strategy to scale network size becomes challenging. In contrast, a quantum router dynamically switches the transmission path of quantum information carriers to configure end-to-end connections over a given network geometry. This can be a more flexible and efficient tool for building up the optimal paths at the quantum information packet level.

Packet-based quantum networking, namely quantum wrapper networking, has been proposed as a paradigm encompassing quantum information with classical control information [10,11,22–25]. Intermediate nodes read the classical header carrying the destination information to route the overall packet without disturbing the quantum payload [10]. Previously demonstrated quantum routers measured classical header information by a photodiode at each routing node and regenerated them for the following nodes [22–24]. Therefore, laser sources are placed at each node with emission wavelength and modulation frequency being precisely locked to the system standards. Our proposal aims to devise a more transparently operating quantum router structure that minimizes insertion loss, background noise, and per-hop latency due to the use of active media and electronic processing.

This work experimentally demonstrates a novel quantum router using partially reflecting optical fiber Bragg gratings (FBGs) that can route the classical and quantum signals successively through multiple hops and nodes by retaining the majority of a classical header except for partial tapping. Classical and quantum data are selectively accessed by a structure analogous to a wavelength-division optical add-drop multiplexer used in optical communications. This implementation helps seamless applicability to existing fiber-based telecommunication networks in real-environment infrastructures. We integrate quantum routers with time-bin-encoded QKD modules using a decoy-state BB84 protocol to realize, for the first time to our knowledge, multi-user packet-wise QKD based on multi-hop quantum routing. One sender selectively sends QKD data to three receivers through double-hop routing by two cascaded quantum routers. Successful measurements of sifted key rates and quantum bit error rates (QBERs) verify the practical usefulness of quantum routing operation for QKD packet delivery.

## 2. Operation Principle

The schematic of our quantum router is shown in Fig. 1. The classical header and the quantum payload of a hybrid packet are separated both by wavelength and time, propagating through different WDM channels. The quantum router partially taps the classical header by reflection of an FBG with reflectivity of 10%, and the reflected header signals are read by a photodiode (PD) via an optical circulator as shown in Fig. 1. The remaining non-reflected portion (90%) of the classical header and the undisturbed quantum payload enter an optical switch (OS; Agiltron NanoSpeed) after a telecom single-mode fiber (SMF) delay of 5 μs (length 1 km). A field-programmable gate array board (FPGA; Digilent Basys3) acquires the classical pulse sequences from the photodiode and actively designates the connection status of the OS to send the packets to the prescribed destination.

The classical and quantum components are placed in adjacent distinguished time slots to reduce photon-level noise by spontaneous Raman scattering (SRS) of the classical signal onto the quantum channel [25–28]. The separation both in wavelength and time suppresses the noise due to SRS and other nonlinear processes caused by comparatively intense classical headers.

The classical header in the current demonstration consists of 2-Mbps 7-bit frames for a total duration of 3.5 μs including a four-bit preamble (fixed as "1010"), a two-bit destination address, and a one-bit parity check, based on on-off keying non-return-to-zero encoding. We note that this type of quantum routing nodes can be cascaded to realize multiple ($N$) hops of routing leading to maximum $2^N$ end nodes, without resorting to active regeneration of the classical headers as long as the remaining optical power ($\propto 0.9^N$) is above the threshold required for code discrimination. Each routing node handles quantum-classical coexistences almost transparently, and precise power management or temporal arrangement of packets are not required.

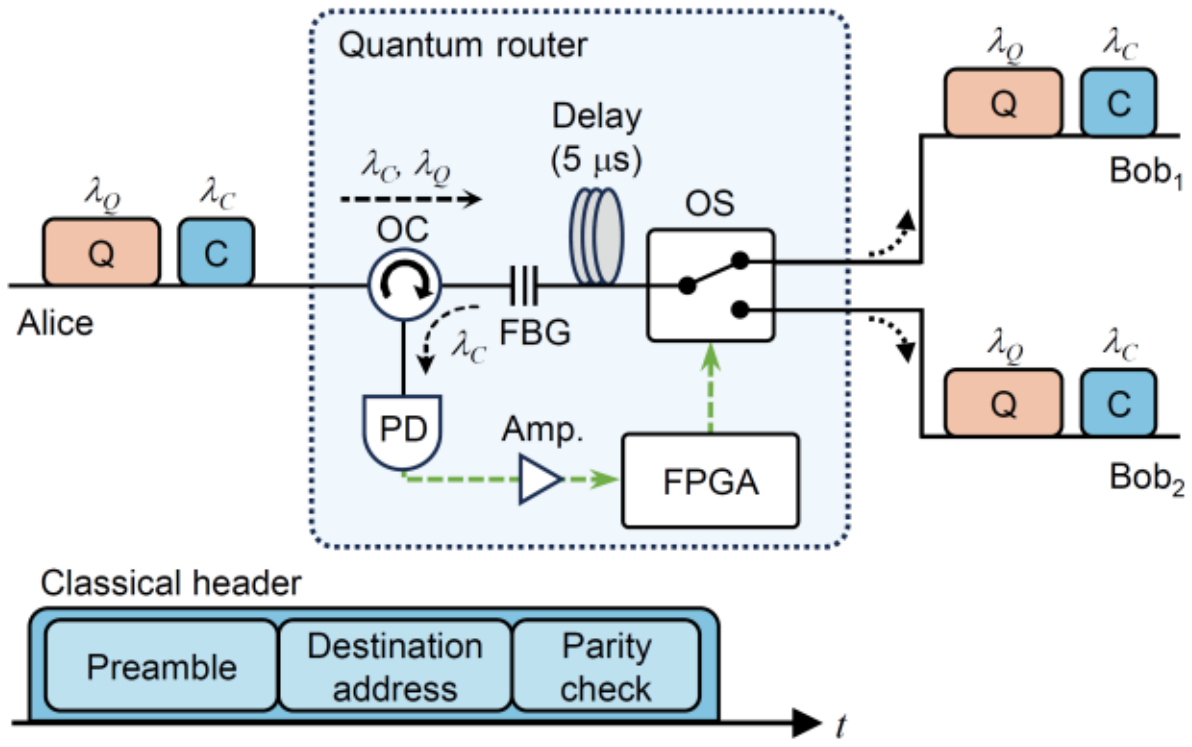

Fig. 1. Schematic of the quantum router and temporal arrangement of a hybrid packet. A classical header has a four-bit preamble, a destination two-bit address, and a one-bit parity check. Alice: sender, Bob$_{1,2}$: receivers, C: classical header, Q: quantum payload, OS: optical switch, FBG: fiber Bragg grating, PD: photodiode, OC: optical circulator, FPGA: field-programmable gate array, Amp.: amplifier.

## 3. Experimental results

### *3.1 Test of the quantum router*

As a natural practical application of the quantum router, we implement a multi-party packet-based QKD system and experimentally demonstrate its operation with three receivers as shown in Fig. 2(a). One sender (Alice) sends hybrid QKD packets and three receivers (Bob$_1$, Bob$_2$, and Bob$_3$) are connected by two cascaded quantum routers (Router$_1$ and Router$_2$). The components were set up within a laboratory, and optical fiber spools with lengths of 20 km, 10 km, and 5 km were inserted before Bob$_1$, Bob$_2$, and Bob$_3$, respectively, to test mid-range communications. The classical header is generated by Alice using a laser diode (LD) with wavelength of 1530.33 nm (ITU-T channel 59) and an electro-optic modulator (EOM). In parallel, the quantum signal is generated using a time-bin-based QKD transmitter with wavelength of 1549.32 nm (ITU-T channel 35). These classical and quantum channels are combined by a dense WDM multiplexer into a common SMF to form a hybrid packet that propagates toward a first quantum router. One of the two output ports was labeled Bob$_1$, and the other output port is connected to the input port of a second quantum router, whose two output ports are labeled Bob$_2$ and Bob$_3$. The classical header generation and the quantum router control are executed on the above-mentioned FPGA, and the quantum signal transmitter (Fig. 2(b)) and receivers (Fig. 2(c)) are controlled by an independent FPGA (NI-9149) sharing only the triggering signals. The routing process is initiated by arrival of the preamble signals by the continually monitoring classical-control FPGA. By arrangement of the optical components, each classical header is placed a guard time [23] of 0.5 μs ahead of the corresponding quantum signal in the combined common optical fiber path.

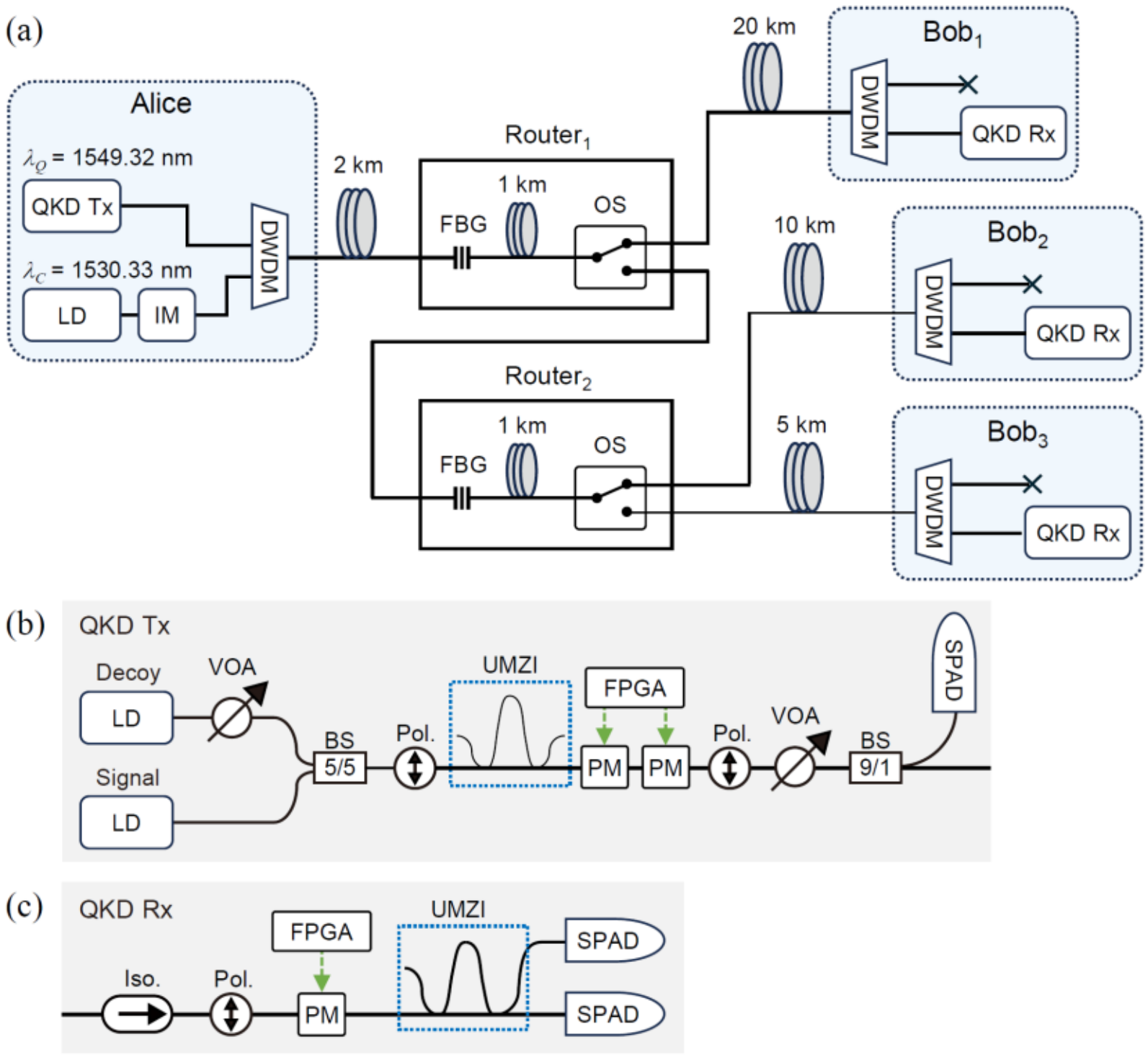

Fig. 2. (a) Experimental setup for quantum routing of time-bin-encoded QKD packets. (b) Structure of the QKD transmitter (Tx). (c) Structure of the QKD receiver (Rx). LD: laser diode, OS: optical switch, IM: intensity modulator, SPAD: single-photon avalanche diode, FBG: fiber Bragg grating, BS: fiber beam splitter, PM: phase modulator, VOA: variable optical attenuator, UMZI: unbalanced Mach-Zehnder interferometer, DWDM: dense wavelength-division multiplexer, Iso.: isolator, Pol.: polarizer, FPGA: field-programmable gate array, ×: termination.

We first test the dynamic switching capability using simple uniform pulse trains as quantum payloads. We cycled through the three packet destinations and measured instantaneous photon counts. Throughout the experiments described in this paper, the destination address is encoded as two-bit non-return-to-zero pulses ($Bob_1$: '00', $Bob_2$: '01', $Bob_3$: '10'). Hybrid packets of a total duration of 1 ms were transmitted with a 1-ms gap between consecutive packets. The first, second, and third hybrid packets were assigned to $Bob_1$, $Bob_2$, and $Bob_3$, respectively, and this sequence was repeated. Each quantum payload was a series of weak coherent pulses (WCPs) with average photon number of 0.5 per pulse, pulse width of 0.5 ns, repetition rate of 10 MHz, and a duration of 1 ms. Single-photon count rates measured at the three receivers are shown in Fig. 3, where the routing steps appear as the time differences of major peaks. The peak power of the classical header signals was set as 1.8 mW considering the photodiodes and amplifiers used to drive the FPGA within the quantum router. Photons were collected with gated-mode InGaAs single-photon avalanche diodes (SPADs) with gating time of 5 ns, and nominal detection efficiency and deadtime were set as 30% and 0.5 μs, respectively. Photon detection events were recorded by a time tagger. The sequentially appearing peaks shown in Fig. 3 are the numbers of photons per 0.1 ms, and verify the successful routing of the packets by the two cascaded quantum routers.

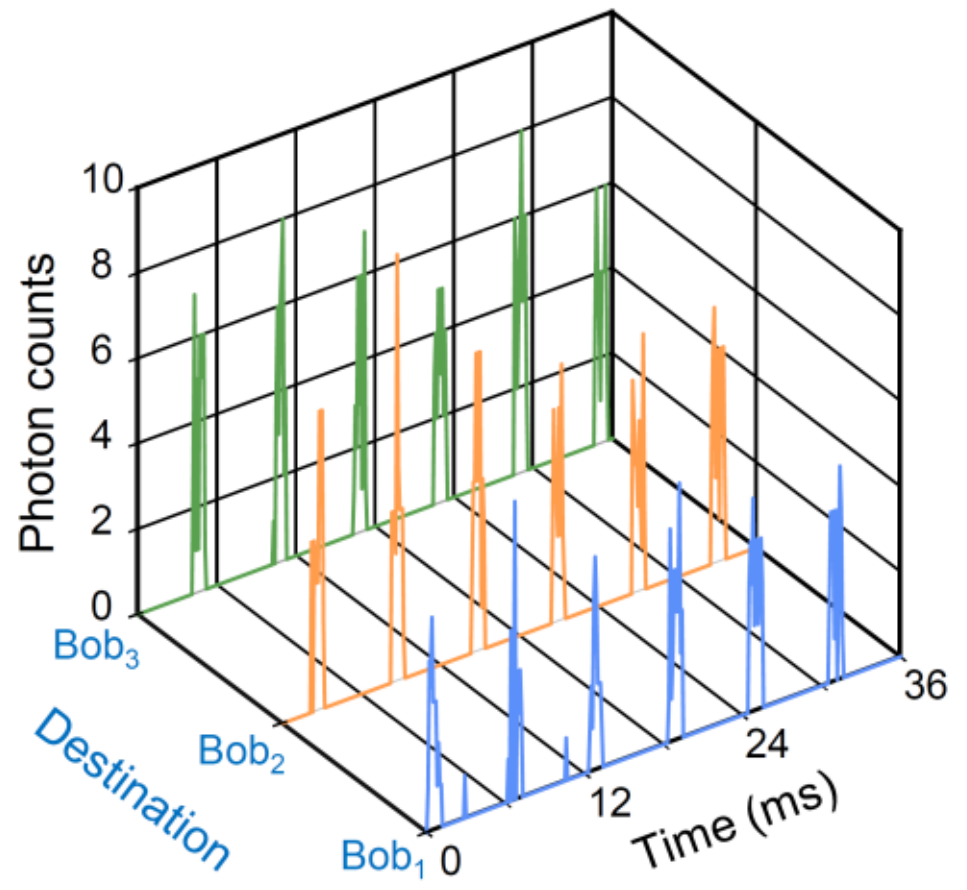

Fig. 3. Photon counts per 0.1 ms at the three receiver nodes with the router states being cyclically changed with period of 2 ms (packet duration: 1 ms, pulse repetition: 10 MHz, average photon number per pulse: 0.5). The observed levels of photon counts reflect the end-to-end transmission from Alice to $Bob_1$, $Bob_2$, and $Bob_3$ of -9.9 dB, -9.5 dB, and -7.9 dB, respectively. Small peaks between major peaks in the $Bob_1$ trace are due to dark counts of a SPAD.

The switching time, a critical metric for the dynamic response of the device, is defined as the minimum transition time between output ports triggered by new arrival of a different address header. The minimum achievable switching time of the current configuration is ~500 ns limited by the rise and fall times of the OSs. The header duration also adds to the processing time, and the current value (3.5 μs) can be further reduced by using faster optical modulation and decoding if necessary. To guarantee a sufficient margin for transition time, incoming hybrid packets experienced the previously mentioned time delay of 5 μs by a buffering SMF spool before entering an OS. The latencies of the other optical and electronic components were negligible compared to the switching time. SRS from the classical-channel wavelength to the quantum-channel wavelength generates significant background noise floor if they can be detected by the SPADs within a common time window [25–28]. Although this effect was suppressed by the 5-ns gating and the 0.5-μs guard time between the classical and quantum signals in the experiments in Fig. 3, imperfect extinction (20 dB) of the OSs can introduce temporal overlap of a classical header ideally addressed to the other nodes and a quantum payload. Therefore, placing a sufficient amount of propagation length differences between receivers are useful to eliminate the influences. The average loss of the quantum signals through the quantum router in Fig. 1 was 2.8 dB, accounting for the insertion losses of an optical circulator (0.5 dB), an FBG (0.1 dB), an OS (1.4 dB), an optical delay line (0.3 dB), and other connector/splicing losses (0.5 dB). The current length of the optical fiber delay did not introduce notable chromatic dispersion into WCPs, however, for quantum networking applications involving heralded or entangled photons with broader bandwidths, it will be necessary to insert dispersion-compensating fibers to eliminate the chromatic dispersion of the SMF delay lines [29].

### *3.2 Delivery of QKD packets to three receivers*

The QKD system was based on time-bin qubits and the BB84 protocol with a two-decoy-state method (weak decoy and vacuum state). Experimental details in terms of qubit structure, operation procedures, and a secret key rate (SKR) estimation method follow our previous literature [30], except that the current transmitter employs an additional second laser to generate weak decoy states while one signal laser was used in [30]. The signal and decoy lasers were matched in center wavelength, spectral width, and pulse width to make the signal and decoy states indistinguishable. During the QKD packet delivery experiments, each QKD packet following a classical header is composed of $10^7$ time-bin qubits (total duration 1 s). WCPs at the quantum channel are first generated by direct modulation of the LD to have pulse width of 0.5 ns and repetition rate of 10 MHz. A time-bin qubit is encoded onto each pulse via a phase-stabilized unbalanced Mach-Zehnder interferometer (UMZI) inscribed on a silica-based planar lightwave circuit, providing a differential optical delay $\tau = 2$ ns between the earlier and later time slots. The relative phase $\phi$ between $|0\rangle$ and $|1\rangle$ states corresponding to earlier and later pulses, respectively, is chosen from $\{0, \pi/2, \pi, 3\pi/2\}$ by the sender using electro-optic phase modulators (PMs) to constitute the four BB84 states. Each receiver detects the QKD qubits along $X$ basis ($\{0, \pi\}$) or $Y$ basis ($\{\pi/2, 3\pi/2\}$) by two gated-mode SPADs by randomly setting the phase of the PM shown in Fig. 2(c) as 0 or $\pi/2$, respectively. The UMZIs at both the sender and receiver nodes were phase-stabilized by temperature control over the housings, by ensuring that photons reach the SPAD that detects bit value 0 in the $X$ basis when phase zero ($\phi = 0$) is applied to all the PMs in the sender and receiver nodes.

To implement the decoy states, the intensity of each pulse was randomly modulated to form a signal, a decoy, or vacuum with average photon numbers of $\mu_s = 0.468$, $\mu_d = 0.186$, and zero, respectively. The relative probabilities of signal, decoy, and vacuum states were 0.66, 0.22, and 0.12, respectively. Each receiver collects sifted key bits from quantum packets, and post-processes the data to estimate final SKRs.

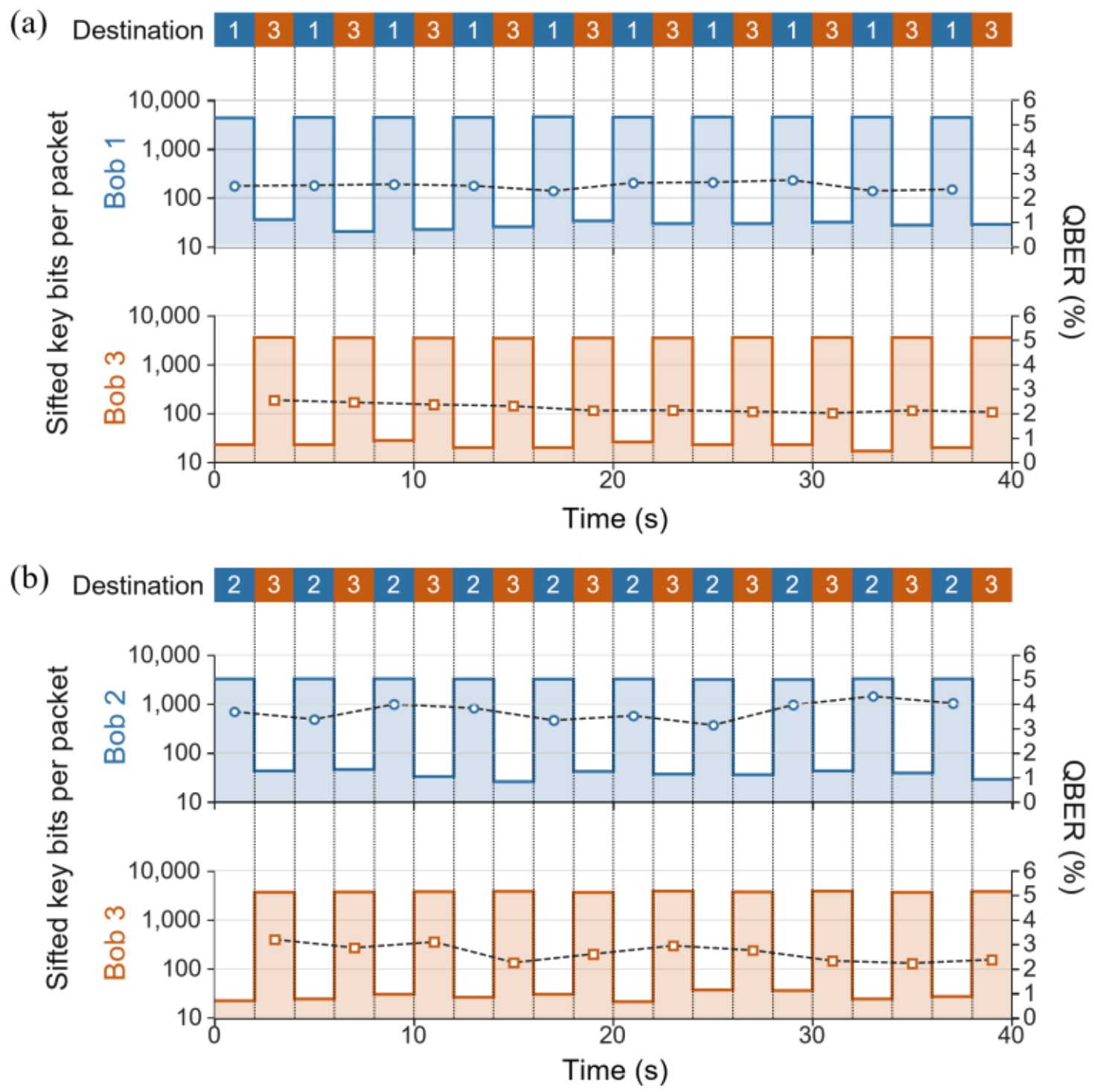

Fig. 4. Sifted key rates (bars) and QBERs (dots) between one sender, Alice, and two receivers: (a) $Bob_1$ and $Bob_3$; (b) $Bob_2$ and $Bob_3$. Time-bin-encoded two-decoy-state QKD packets with duration of 1 s and pulse repetition rate of 10 MHz were alternately delivered to the two receivers by the quantum routers.

Figure 4 shows the sifted key rates for the cases that Alice sends hybrid packets with two alternating destination addresses among the three receivers. Sifted keys of 4–5 kbits on average were detected for each packet of 1 s between Alice and the proper receivers as shown in Fig. 4. The average QBER over the three receivers was 3.4%, which is expected given the imperfect extinction of the classical header (22 dB), the dark count probability of SPADs ($2\times10^{-5}$), and non-zero crosstalk of phase modulation between the two time-bins (~17 dB). As an indicator of relative degradation due to the quantum routing, the difference of QBER with the classical-channel laser being turned on and off was measured to be 0.14%. We attribute this error to SRS by the leaked classical light that was incompletely suppressed by electro-optic amplitude modulation during header signal encoding that had the finite extinction ratio of 22 dB.

SKRs from the acquired sifted keys were extracted by the following procedures. In the finite-key regime, the secret key length after classical post-processing depends on the block size. We set a block size of $1.5\times10^6$ bits in this experiment. Winnow information reconciliation and the privacy amplification using Toeplitz matrices were applied to the acquired sifted keys. The SKRs were calculated by dividing the resulting secret key length by the corresponding acquisition periods. The secrecy parameter $\varepsilon_{sec}$ and the secret key length $l = RT$, where $R$ and $T$ are the SKR and the total acquisition duration, respectively, are calculated according to the equation [31]:

$$l = s_{X,0} + s_{X,1} - s_{X,1}h(\phi_X) - \lambda_{EC} - 6\log_2\frac{21}{\varepsilon_{sec}} - \log_2\frac{2}{\varepsilon_{cor}}, \quad (1)$$

where $h(\phi_X)$ denotes the binary Shannon entropy, and $s_{X,0}$ and $s_{X,1}$ are the numbers of vacuum and single-photon events, respectively, in the $X$ basis. The parameter $\phi_X$ is the phase error rate associated with the single-photon events in the $X$ basis, and $\lambda_{EC}$ represents the amount of information leaked during the error correction process. The last two terms account for the finite-key corrections associated with secrecy and correctness, respectively. To obtain the expected secret key length, we used a security constant $\kappa = 10^{-13}$ and a protocol correctness parameter $\varepsilon_{cor} = 10^{-15}$. The secrecy parameter $\varepsilon_{sec}$ is set proportional to the secret key length as $\varepsilon_{sec} = l \times \kappa$, where $\kappa$ represents the secrecy leakage per generated bit. $\varepsilon_{sec}$ thus depends on $l$, and Eq. (1) is solved self-consistently for $l$. Calculation based on the above equation resulted in non-zero experimental SKRs of 35–270 bps over one-hop ($Bob_1$) and two-hop ($Bob_2$ and $Bob_3$) routing cases, as shown in Fig. 5. The theoretical SKR curves were calculated based on the experimental data collected by $Bob_1$, and the average transmission loss (2.8 dB per node) of the quantum routers were applied to draw 0-, 2-, and 3-router cases. The causes of relatively low SKRs of $Bob_2$ are currently unclear and believed to reflect the instability of current experimental setup, such as birefringence of optical fiber spools and ambient temperature, during the whole period (1.6 hour) of key collection.

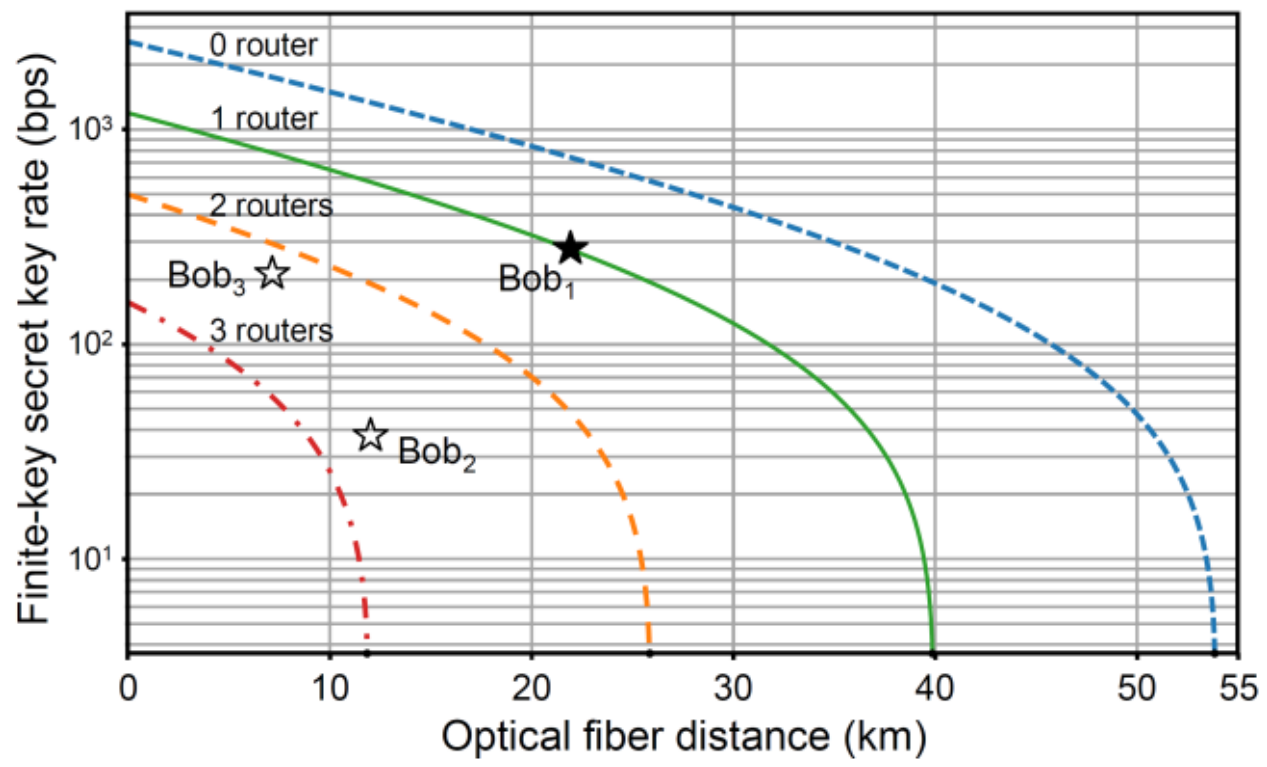

Fig. 5. Experimentally extracted secret key rates (stars) and theoretical fits based on sifted key rates and QBERs of $Bob_1$ (lines). The fiber delays within the quantum routers are considered only as loss elements and not added to the optical fiber distance. Both the experimental SKRs and theoretical fits used $\kappa = 10^{-13}$, $\varepsilon_{cor} = 10^{-15}$, and a finite-key block size of $1.5\times10^6$ bits. The error-correction efficiency was determined for each experimental block from the Winnow reconciliation results, and a fixed value of 1.39 obtained for $Bob_1$ was used for all the theoretical fits.

## 4. Discussion and conclusion

In summary, we have proposed and demonstrated a novel quantum router based on optical fiber devices and double-hop routing of hybrid quantum-classical packets onto three receivers. The experimental results verify the operation principles offering advantages of maintaining quantum information intact and accessing the classical address information multiple times without regeneration. The QBER level of the time-bin-encoded QKD signals was kept below threshold for secret key generation, although possibly high in a practical point of view due to experimental imperfections. Primarily, we expect that the error rate can be further reduced by improving the extinction ratio of classical-signal generators and mitigating unwanted crosstalk between time-bins during the preparation and measurement of qubits. The noise in the quantum channel caused by the classical headers was estimated by comparing the QBERs with turning on and off the classical-channel light source, and the resulting difference was 0.14%.

The key performance of the proposed quantum router is currently governed by the operation speed (~ 5 μs) and transmission efficiency (≤−1.5 dB) of optical switches. Unlike the previous quantum router structures, quantum payloads do not propagate through time- or wavelength-selective devices that can cause additional losses. The sequential attenuation of the classical channel's signal intensity due to tapping (10%) in each node can be compensated for by optical amplifiers placed at relevant intervals. Therefore, the level of transmission efficiency for quantum channels eventually determines the coverage of quantum router-assisted network in terms of the number of nodes and the transmission distance. Low-loss and high-speed optical switching technology will thereby enhance these characteristics while also advancing various photonic quantum information processing technologies. For example, all optical switches based on the Kerr effect have been demonstrated with losses as low as 2.3 dB and switching times on the picosecond scale. [32]

Future work may extend to demonstration of advanced quantum communication protocols over long-distance multi-node quantum networks supporting distribution of entangled photons. This will require technical components such as synchronization and birefringence stabilization, and it is noteworthy that the classical header part of hybrid packets can be used for these purposes [33]. We expect that compatibility of the quantum router method with protocols utilizing phase stability over a finite timescale, such as mode-pairing QKD [34], also deserves further investigation. The successful deployment of the active quantum routing capability adds an important milestone in transitioning from point-to-point quantum links to scalable, multi-user quantum networks, and serves as a groundwork for the future quantum internet.

**Funding.** National Research Foundation of Korea (RS-2022-NR068817, RS-2023-00256050); National Research Council of Science & Technology (GTL25011-000, CAP22054-000); KRISS (GP2026-0010-04); KRISS-KAIST quantum graduate program.

**Acknowledgment.** T. Y. Park and H. S. Park thank Dr. H. J. Lim and Dr. S. J. Ha for discussions.